\documentclass[aps,twocolumn,superscriptaddress,prb,amsmath,amssymb,showpacs,lengthcheck,reprint,floatfix]{revtex4-1} %revtex4-1
\usepackage[T1]{fontenc}
\usepackage[latin9]{inputenc}
\usepackage{mathrsfs}
\usepackage{graphicx}
\usepackage[usenames,dvipsnames]{color}

       \definecolor{mygray}{gray}{0.88}
     \newcommand{\note}[1]{{ #1 }}
     \newcommand{\noteblue}[1]{{ #1 }}
\usepackage{braket}
\usepackage{placeins}

\usepackage{blindtext}

\begin{document}

\title{Strategies for the alignment of electronic states in quantum-dot tunnel-injection lasers }  
%and their influence on the emission dynamics
\author{Michael Lorke}
\email{mlorke@itp.uni-bremen.de}
\affiliation{Institute for Theoretical Physics and Bremen Center for Computational Materials Science, University of Bremen, P.O. Box 330 440,  Bremen, 28334,  Germany}
\author{Igor Khanonkin}
\affiliation{Electrical Engineering Department and Russel Berrie Nanotechnology Institute, Technion, Haifa, 32000, Israel}
\author{Stephan Michael}
\affiliation{Institute for Theoretical Physics and Bremen Center for Computational Materials Science, University of Bremen, P.O. Box 330 440,  Bremen, 28334,  Germany}
\author{Johann Peter Reithmaier}
\affiliation{Technische Physik, Institute of Nanostructure Technologies and Analytics, Center of Interdisciplinary Nanostructure Science and Technology (CINSaT), University of Kassel, Kassel, 34132, Germany}
\author{Gadi Eisenstein}
\affiliation{Electrical Engineering Department and Russel Berrie Nanotechnology Institute, Technion, Haifa, 32000, Israel}
\author{Frank Jahnke}
\affiliation{Institute for Theoretical Physics and Bremen Center for Computational Materials Science, University of Bremen, P.O. Box 330 440, Bremen, 28334, Germany}

\begin{abstract}
In quantum-dot tunnel-injection lasers, the excited charge carriers are efficiently captured from the bulk states via an injector quantum well and then transferred into the quantum dots via a tunnel barrier. The alignment of the electronic levels is crucial for the high efficiency of these processes and especially for the fast modulation dynamics of these lasers. In particular, the quantum mechanical nature of the tunneling process must be taken into account in the transition from two-dimensional quantum well states to zero-dimensional quantum dot states. This results in hybrid states, from which the scattering into the quantum-dot ground states takes place. We combine electronic state calculations of the tunnel-injection structures with many-body calculations of the scattering processes and insert this into a complete laser simulator. This allows us to study the influence of the structural design and the resulting electronic states as well as limitations due to inhomogeneous quantum-dot distributions. We find that the optimal electronic state alignment deviates from a simple picture in which the of the quantum-dot ground state energies are one LO-phonon energy below the injector quantum well ground state.
\end{abstract}
\maketitle

%\section{Introduction}
Tunnel-injection (TI) lasers with quantum dots (QD) as an active material promise improved emission properties, in particular faster modulation speed in optical data transmission networks. \cite{Sun:93,Bhattacharya:92,Han:08,Bhowmick:14,gready2010carrier} 
In conventional QD lasers with current injection pumping, the pump process generates excited carriers in the bulk barrier states. Capturing these carriers into the QD states requires substantial energy dissipation via carrier-phonon-scattering in combination with carrier-carrier scattering. This limits the excitation efficiency for electrons and holes in the QD states and slows down the modulation speed of these lasers. The TI-laser design utilizes an injector quantum well (QW) that is directly coupled to the QDs via a tunnel barrier. The injector QW provides a much larger scattering cross section to the excited carriers driven by the carrier injection, thereby enhancing the capture rate of carrier from the bulk states. Tunnel coupling facilitates efficient transfer of excited carriers from the injector QW into the QD states.

The design of TI-QD lasers requires proper alignment of the electronic states between the injector QW and the QDs in order to exploit the advantages of this system. The dominant carrier scattering process, which contributes to the tunnel-assisted carrier transfer from the injector QW to the QDs has been identified as carrier interaction with LO phonons.  \cite{Sun:93,Bhattacharya:92,Bhowmick:14,gready2010carrier,michael2018interplay,lorke2018performance,Lorke2022carrier} Hence the original design goal was to engineer the structure in a way that the QD ground states are positioned one LO-phonon energy below the band edge of the injector QW.

A closer look reveals two challenges. Due to an inhomogeneous distribution of QD confinement energies, only some of the QDs have optimal tunnel coupling. In the presently used material systems like InGaAs, the inhomogeneous broadening of energies has a FWHM of 15 to 25 meV in comparison to the LO-phonon energy of 36 meV. In addition, tunnel coupling into the QDs is inefficient for the lowest conduction band states of the injector QW. Optimal tunnel coupling for the conduction-band states is linked to hybrid states between the first excited QD states and injector QW states with finite carrier momentum. The last two statements have been derived from a quantum mechanical treatment of the electronic coupling between the 2d and 0d single-particle states, \cite{michael2018interplay,lorke2018performance,Lorke2022carrier}
which leads to hybrid states that are partly localized in both the injector QW as well as in the QDs. The transfer of excited charge carriers from the QW into the QD then takes place in the form of scattering processes between these hybrid states and the QD ground state.

We have developed a microscopic description for the excited charge carrier transfer from the injector QW to an inhomogeneously distributed QD ensemble. Calculations of the electronic states in the coupled 2d-0d system serve as a starting point. These provide single-particle states and interaction matrix elements for the carrier-phonon interaction and carrier-carrier Coulomb interaction. While Boltzmann-type scattering rates in kinetic equations could have been used as a subsequent step, we find that a quantum kinetic calculation of scattering rates provide a more appropriate description, as it accounts for polaron energy renormalizations and deviations from a strict energy conservation in terms of free-particle states. \cite{Seebeck:05,michael2018interplay,lorke2018performance,Lorke2022carrier} The results determine capture rates of excited electrons into the QDs as function of their detuning and the resulting carrier dynamics.
Here the carrier scattering through emission of LO phonons has been identified as the dominant process while carrier-carrier Coulomb scattering additionally contributes.

\begin{figure}[h!]
  \begin{center}
%    \hspace*{2cm}%
    \includegraphics[width=0.5\textwidth,angle=0]{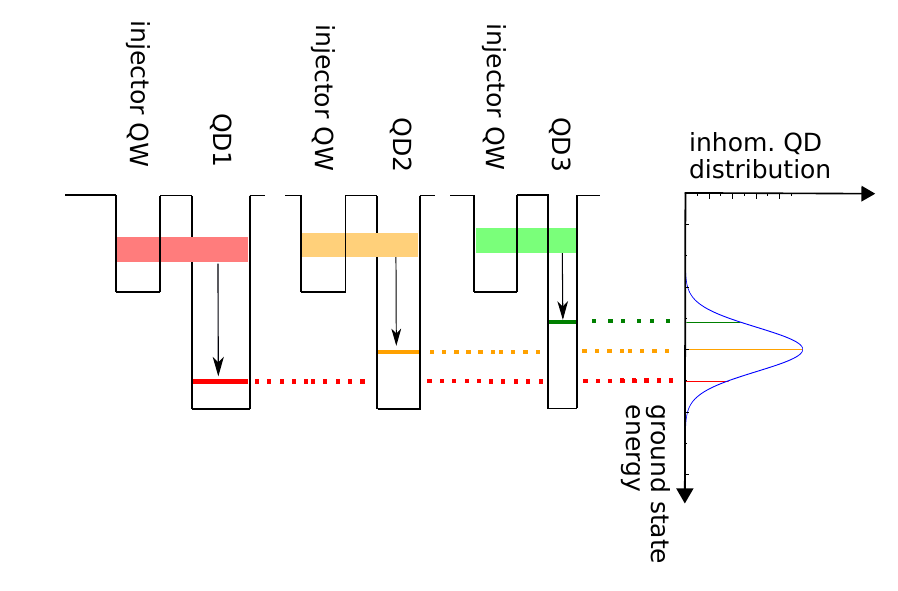}
    \caption{
Inhomogeneous distribution of QD ground state energies (blue line) and hybridized p-shell states within the injector QW continuum (shaded area). Top: Examples for three different QDs with zero-dimensional ground state (s-shell). 
These QDs do not have a localized p-shell, as these states hybridize with the injector QW continuum states. 
    \label{schematic}}
  \end{center}
\end{figure}

In this paper we integrate our description of the carrier kinetics in tunnel-coupled injector QW and QD system into a complete laser simulator. The latter includes a spatially resolved modeling of the carrier transport in a structure with 6 injector QWs and the coupling of the excited carrier recombination in the QDs via spontaneous and stimulated emission to a self-consistent laser model. We consider a QD ensemble with size and composition fluctuations leading to a Gaussian distribution of ground state (s-shell) energies with a FWHM of 15meV. The maximum of the distribution is centered at an interband transition energy of 800meV (1550nm). The situation is sketched in Fig.~\ref{schematic} with the blue line. 
In addition three examples for QDs in the inhomogenoues ensemble are shown with their ground state energy (solid colored line) and the associated hybridized injector QW-QD states (colored shaded areas). The latter are the new eigenstates due to tunnel coupling, which are formed from zero-dimensional excited QD states and two-dimensional injector QW states. For a given QD, the hybridization takes place for a range of QW energies. The energy with strongest coupling is lowered in comparison to the uncoupled excited QD state, as the tunnel coupling weakens the confinement of this state. The hybridized states are calculated for the individual QDs and vary to some extent within the inhomogeneous ensemble. This applies in particular to the wave function symmetry when involving low carrier momentum injector QW states.
\cite{Lorke2022carrier}

In this work, we compare the previously suggested TI design, in which the energetic distance between the QD ground state and the injector QW ground state
is tuned to the LO phonon energy (alignment A in what follows), with a situation that we propose in this work,
where the injector QW is shifted to lower energies in an effort to optimize the scattering efficiency (alignment B).
To study switch-on and modulation properties of lasers with TI structures as active material,
we use a model that includes both the microscopic carrier dynamics in the TI structure,
as well as the carrier transport through the bulk material.

The charge carrier transport in the bulk material regions, including the injection of carriers and their distribution through the structure via drift and diffusion is described on a classical level by
\begin{equation}
\begin{split}
\frac{d}{dt}N_{\text{res}}&(z,t)|_\text{transport} \\
& =\frac{d}{dz}\left[D_N \frac{d}{dz} N_{\text{res}}(z,t)+\mu_N N_{\text{res}}E(z,t)\right]
\end{split}
\end{equation}
with boundary condition
\begin{equation}
J_N=q\left[D_N \frac{d}{dz} N_\text{res}(0,t)+\mu_N N_\text{res}E(0,t) \right]~.
\end{equation}
In this equation $N_{\text{res}}$ is the three-dimensional carrier density of the bulk material, $D_N$
and $\mu_N$ are the diffusion coefficient and mobility of the carriers and $E(z,t)$ is the internal
field obtained from a solution of Poisson's equation for the carrier density. 

The net result for the carrier dynamic in the bulk material contains coupling to the active region, as well as a loss term for the reservoir \note{due to spontaneous recombination with the time constant $\tau_\text{res}$},
\begin{equation}
\begin{split}\label{Nres}
\frac{d}{dt}N_{\text{res}}(z,t)
&=\frac{d}{dt}N_{\text{res}}(z,t)|_\text{transport}-\frac{N_{\text{res}}(z,t)}{\tau_\text{res}}\\
&-N_{\text{res}}(z,t)\sum_{i=1}^M \frac{N_\text{QD}^i}{\tau^i_\text{cap}} (1-f^{i;e,h})\\
&+\sum_{i=1}^M\frac{N_{QD}^i}{\tau^i_\text{esc}}f^{i;e,h}\left(1-\frac{N_{\text{res}}}{D_\text{res}}\right)\\
&-\sum_{j}\frac{N_{\text{res}}}{\tau_\text{cap,QW}}\left(1-\frac{N^j_\text{QW}}{D_\text{QW}}\right)\\
&+\sum_{j}\frac{N^j_\text{QW}}{\tau_\text{esc,QW}}\left(1-\frac{N_{\text{res}}}{D_\text{res}}\right)~.
\end{split}
\end{equation}
The second and third lines of Eq.~\eqref{Nres} describe direct scattering of the carriers from/to the bulk to/from the QD states with times constants $\tau^i_\text{cap}$ and $\tau^i_\text{esc}$, respectively, where  $f^{i;e,h}$ the electron and hole occupation of the of the QD s-shell. 
\note{Originally, the sum over $i$ numbers the QDs with inhomogeneous distribution of transition energies. For practical purposes, we group QDs with similar electronic properties, where $M$ is the number of sub-ensembles and $N_\text{QD}^i$ is the number of QDs in the sub-ensemble.}

The fourth and fifth rows of Eq.~\eqref{Nres} describe capture and escape form the bulk states into the injector QW with carrier density $N^j_\text{QW}$ and density of states $D_\text{QW}$. The respective time constants are $\tau_\text{cap,QW}$ and $\tau_\text{esc,QW}$. The index j numbers the different injector QWs.

The equation of motion for the injector QW density follows as
\begin{equation}
\begin{split}
\frac{d}{dt} N^j_\text{QW}&=-\frac{N^j_\text{QW}}{\tau_\text{QW}}\\
&-\sum_{i=1}^M \frac{N^j_\text{QW}}{\tau^i_\text{tun}}N_\text{QD}^i (1-f^{i;e,h}) \\
&+\sum_{i=1}^M\left(1-\frac{N^j_\text{QW}}{D_\text{QW}}\right) 2 N_\text{QD}^i\frac{f^{i;e,h}}{\tau^i_\text{tun,b}} \\
&+ \frac{N_{\text{res}}(z^j_\text{QW},t)}{\tau_\text{cap,QW}}\left(1-\frac{N^j_\text{QW}}{D_\text{QW}}\right)\\
&-\frac{N^j_\text{QW}}{\tau_\text{esc,QW}}\left(1-\frac{N_{\text{res}}(z^j_\text{QW},t)}{D_\text{res}}\right)~,
\end{split}
\end{equation}
where the first line \note{represents the loss of carriers due to spontaneous recombination with the time constant $\tau_\text{QW}$}. The second and third line account for carrier scattering between the hybridized injector QW states and the QD s-shell with the associated rates $1/\tau^i_\text{tun}$ and $\tau^i_\text{tun,b}$, as describe in our previous publication \cite{Lorke2022carrier} and in the Supplement.
The fourth and fifth line are the counter terms to the respective contributions in 
Eq.~\eqref{Nres}. Here $z^j_\text{QW}$ is the position of the $j$-th injector QW.

The ground-state population of the $i$-th QD, $f^{i;e,h}$, is calculated from the semiconductor Bloch equations \cite{Haug_Koch:04} with the added contributions due to the carrier scattering as introduced above. The QD population dynamics 
\begin{equation}
\begin{split}
\frac{d}{dt}f^{i;e,h}=&- \frac{f^{i;e}f^{i;h}}{\tau_\text{sp}} -2\text{Im}\left(\Omega_\text{lase}\psi^{i,\ast}\right)\\
&+\frac{N_{\text{res}}(z^i_\text{QD},t)}{2\tau^i_\text{cap}}(1-f^{i;e,h})\\
&-\frac{f^{i;e,h}}{\tau^i_\text{esc}}(1-\frac{N_{\text{res}}(z^i_\text{QD},t)}{D_\text{res}})\\
&+\sum\limits_j\frac{N^j_\text{QW}}{2\tau^i_\text{tun}}(1-f^{i;e,h})\\
&-\sum\limits_j\left(1-\frac{N^j_\text{QW}}{D_\text{QW}}\right)\frac{f^{i;e,h}}{\tau^i_\text{tun,b}} 
\end{split}
\end{equation}
is coupled to the optical interband polarization
\begin{equation}
\frac{d}{dt}\psi^i = - (i\omega+\gamma_h)\psi^i- i\frac{\Omega_\text{lase}}{\hbar}\left( 1-f^{i;e}-f^{i;h}\right)~.
\end{equation}
Here $\tau_\text{sp}$ is the spontaneous emission time and $\gamma_h$ is the homogeneous linewidth.
Furthermore, $\Omega_\text{lase} = d E(z^i_\text{QD},t)$ contains the interband dipole coupling matrix element $d$ as well as the laser field $E$ at the QD position $z^i_\text{QD}$.
The laser field $E(z,t)$ can be written in single-mode approximation as 
\begin{eqnarray}
  E(z,t) & = & \frac{1}{2} \mathscr{E}(t) u(z) e^{-i \omega_c t} + c.c.
\end{eqnarray}
with 
\begin{eqnarray}
  \frac{d\mathscr{E}}{dt} & = & -\gamma_{c} \mathscr{E} + \frac{i\omega_c}{\varepsilon_{b}} \sum_{i=1}^M N_\text{QD}^i d^{*} \psi^{i} + \frac{f^{i;e}f^{i;h}}{\tau_\text{sp}}~,
\end{eqnarray}
where $u(z)$ is the mode function, $\omega_c$ is the laser frequency, and $\gamma_{c}$ describes the cavity field loss rate.

We consider an inhomogeneous QD ensemble with size and composition fluctuations, described by a Gaussian distribution of ground state (s-shell) energies. The QD layers are separated from the injector QW by a 2.1 nm tunnel barrier. The active material of the laser consists of 6 QW-QD layers separated with 10.5 nm wide spacers layers. 

\begin{figure}[h!]
  \begin{center}
%   \hspace*{-2cm}%
    \includegraphics[width=0.5\textwidth,angle=0]{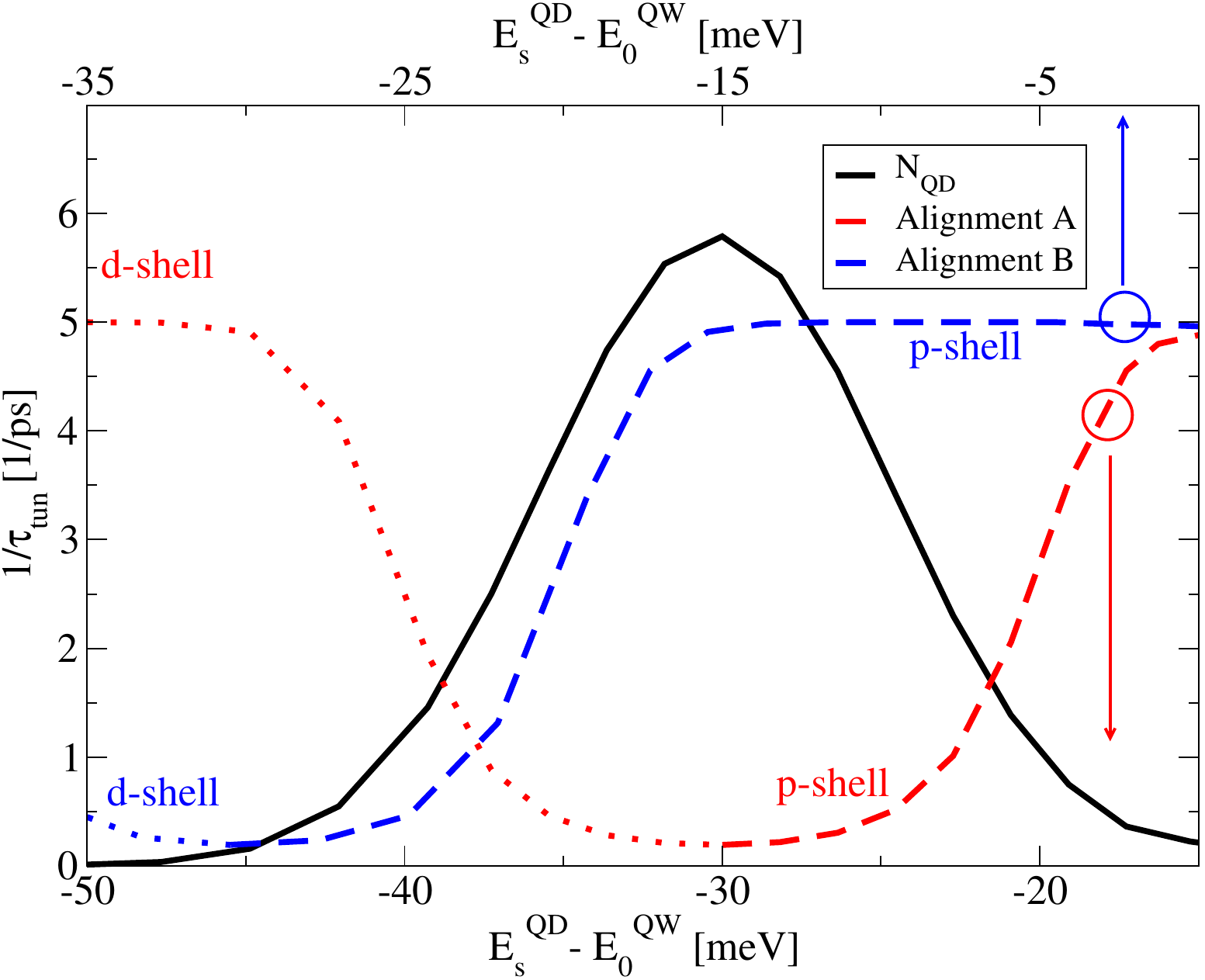}
    \caption{
    Carrier capture rates from the injector QW into the QD as function of the QD ground-state energy relative to the QW zero-momentum state. Dashed lines correspond to the situation in which the p-shell is hybridized, while for the dotted line the p-shell is confined and the d-shell hybridizes with the QW. Red lines (alignment A, original design recommendation) refer to the bottom energy scale and blue lines (alignment B, new design recommendation) to the top energy scale. For both cases the inhomogeneous QD distribution is shown as black line.
    \label{align}}
  \end{center}
\end{figure}

\noteblue{Carrier scattering rates into the QD ground state are calculated by means of a quantum-kinetic theory which includes renormalizations of the electronic energies due to the electron-phonon coupling (polaronic effects \cite{Seebeck:05}).
In this way we obtain an accurate description of carrier scattering by interaction with LO phonons, which according to our studies in Ref.~\cite{michael2018interplay} forms the dominant scattering channel and is even more efficient than carrier-carrier Coulomb scattering, which we also considered there.
We use three-dimensional quantum mechanical calculations of the electronic states in the coupled injector QW - tunnel barrier - quantum dot system using the nextnano$^3$ software package \cite{nextnano3}. The resulting wavefunctions are directly applied in calculations of the interaction matrix elements for the scattering processes. On this basis, the influence of the TI design on the charge carrier dynamics in the laser is considered on a microscopic level.
}
\noteblue{Further details can be found in the supplementary material.}
%Calculated 
\noteblue{The resulting} carrier capture rates into the QD ground state as function of its energy relative to the bottom of the injector QW conduction band are shown in Fig.~\ref{align} together with the QD distribution (black line).
We find that the capture efficiency is small, if the QD s-shell energy is one LO-phonon energy below the bottom of the injector QW. This is a result of the wavefunction symmetry. The capture efficiency steadily increases for higher energies until it drops due to the dereasing carrier population of these states, providing a capture bandwidth of more than 15 to 20 meV. If the detuning of the QD ground state below the bottom of the injector QW is too large (the p-shell is below the band edge of the injector QW), the p-shell can no longer hybridize and the QD d-shell takes over the carrier capture.

The red lines in Fig.~\ref{align} represent the original design proposal for TI-QD lasers. When comparing the capture rate relative to the inhomogeneous QD distribution (black line), the majority of QDs does not benefit from efficient hybridization. For the high-energy tail of the QD distribution, the QD p-shell states are more than 15 meV above the injector QW band edge and hybridize increasingly efficient, thus providing higher capture rates. The low-energy tail of the QD distribution has d-shell energies $\gtrsim$ 15meV above the QW band edge, thus allowing for their efficient hybridization. Reflecting the tail of the carrier population in the injector QW, their capture rate increases for smaller energies.

Our main finding is that the scattering reaches its maximum if the QD ground states couple via LO phonons with 36 meV energy to injector QW states approximately 15 meV above their conduction band edge. This leads to our proposed alignment B (blue lines in Fig.~\ref{align}), which places the injector QW energetically closer to the QD ensemble, thereby enabling improved carrier scattering at the maximum of the QD distribution. This alignment could be achieved, for example, by changing the injector QW width or concentration profile. 
We would like to point out that this effect is mainly the result of wavefunction engineering and can be also verified with a perturbative description of carrier scattering, based on Fermi's golden rule.

\begin{figure}[h!]
  \begin{center}
%   \hspace*{-2cm}%
    \includegraphics[width=0.5\textwidth,angle=0]{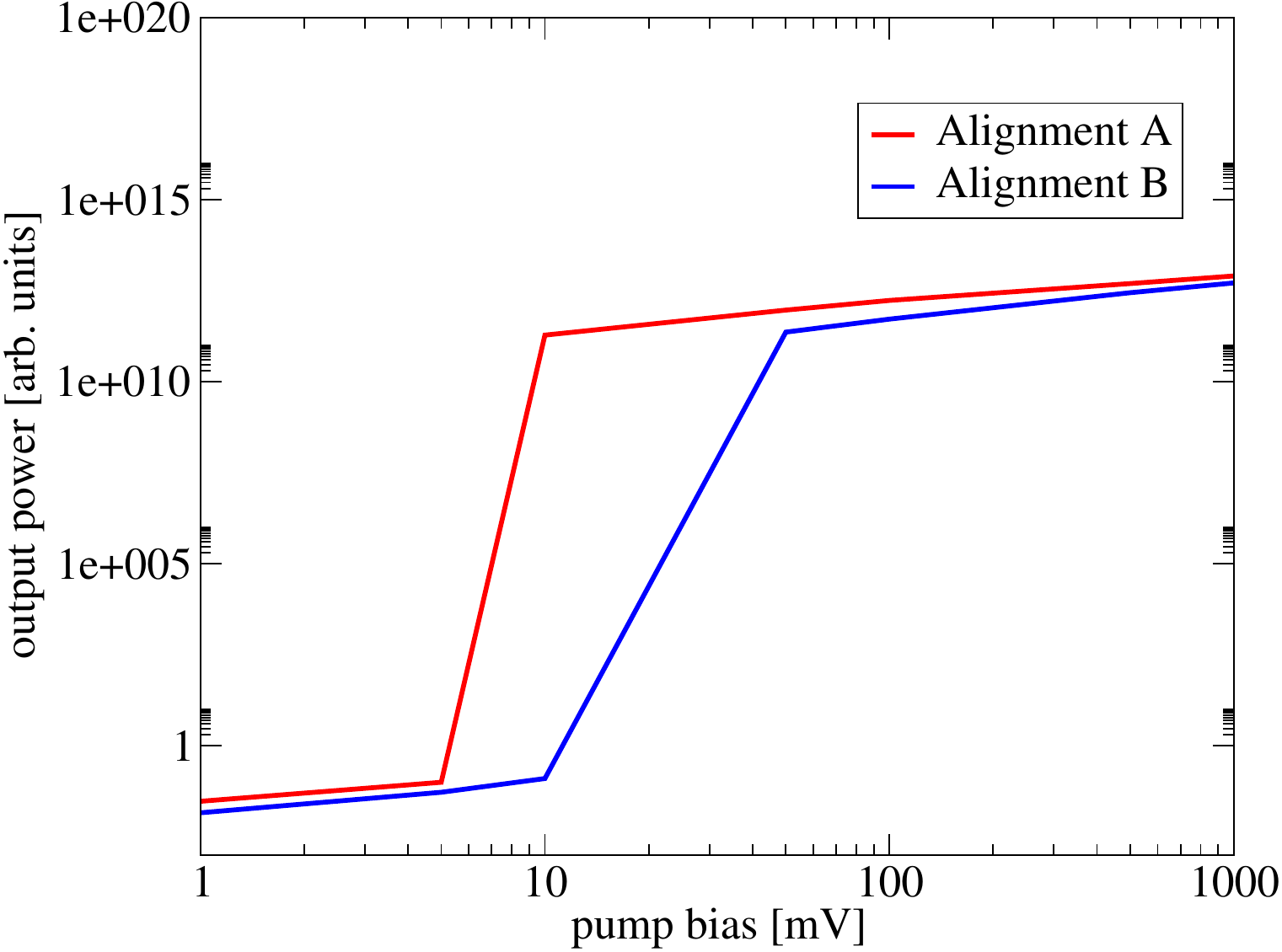}
    \caption{
    Input/Output characteristics for TI laser devices with 6 active layers using the previously recommended (red) and optimized (blue) alignment between the injector QW and the QD ensemble
    \label{io}}
  \end{center}
\end{figure}

\begin{figure}[h!]
	\begin{center}
        \includegraphics[width=0.5\textwidth,angle=0]{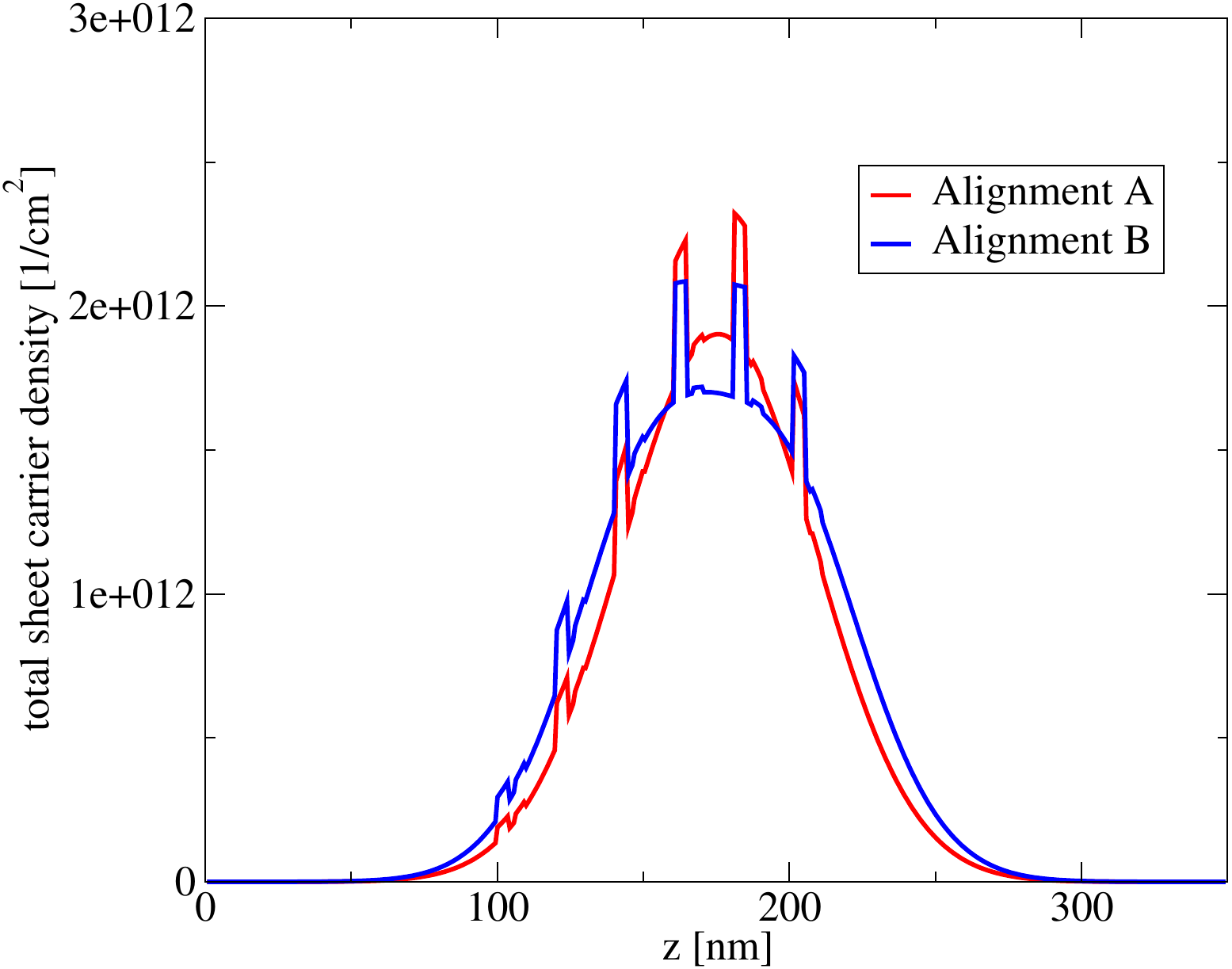}
		\caption{
			\label{zdep} Total electron density as a function of position for the two discussed alignments
of the QD energy distributions relative to the injector QW.}
	\end{center}
\end{figure}

In Fig.~\ref{io} calculated input-output characteristics are shown for alignments A and B. 
It is worth noting that the proposed alignment B has a higher threshold current than the previously suggested alignment A.
This is the result of a more uniform distribution of the carrier population across the 6 active layers with the optimized alignment B (see Fig.~\ref{zdep}), which requires a higher carrier density in the device before lasing can occur. Specifically, additional carriers are need in the outer layers so that they can contribute better to laser emission.

To investigate this further, the total carrier density $(N_\text{res}+N_\text{QW}+N_\text{QD} f^{i;e})$ is
shown in Fig.~\ref{zdep} for a pump power 10\% above the laser threshold for both devices. Electrons (holes) are generated by contacts at edge of the device corresponding to the left (right) side in this figure.
For alignment A, the slower carrier capture rate leads to carriers being more localized 
in the central layers of the device. 

\begin{figure}[h!]
	\begin{center}
		\includegraphics[width=0.5\textwidth,angle=0]{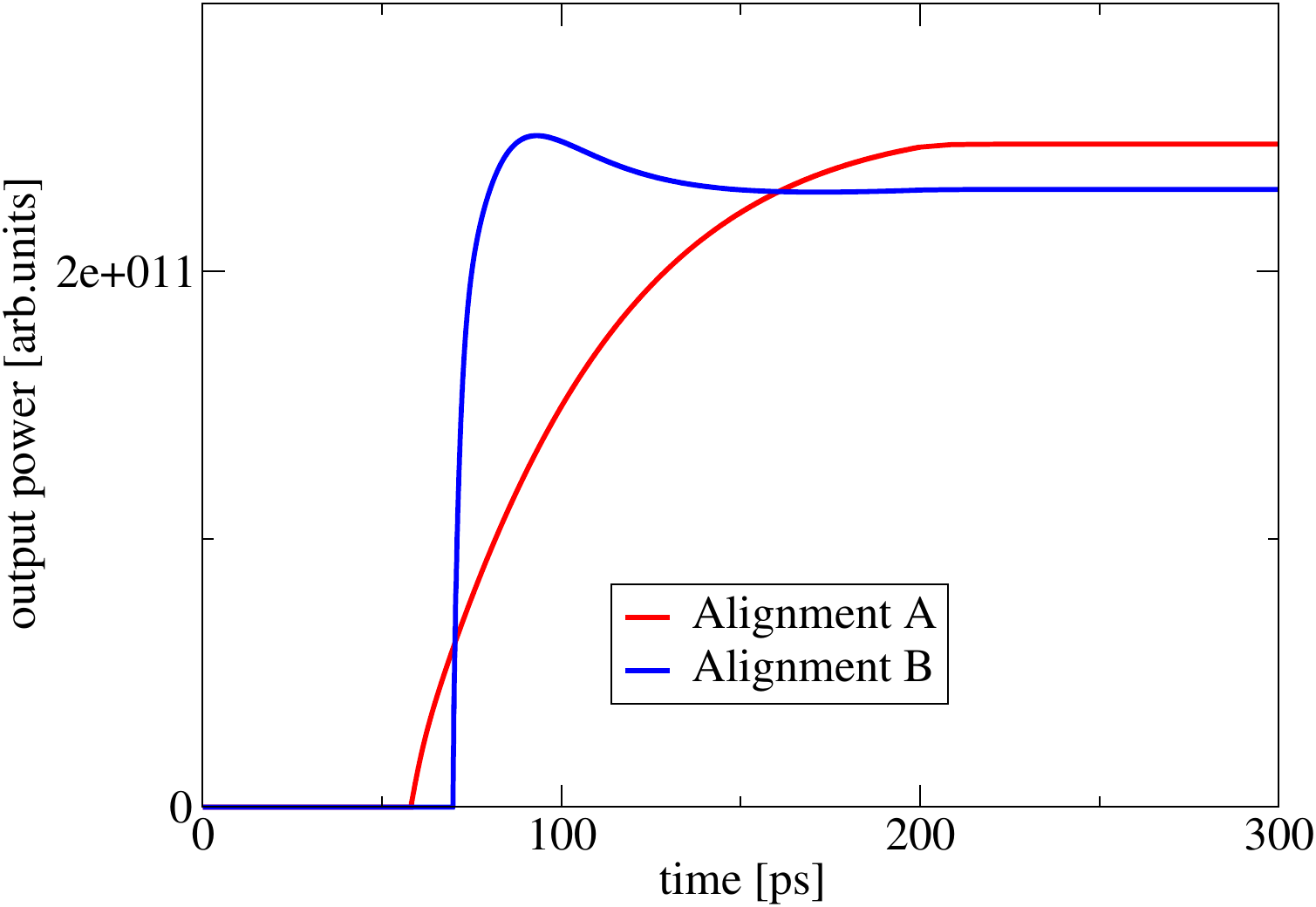}
		\caption{
			\label{switch} Output power as function of time using a pump power 10\% above the laser threshold for the two discussed alignments of the QD energy distributions relative to the injector QW.}
	\end{center}
\end{figure}

While the higher threshold seems to indicate that the proposed "optimized" alignment B performs worse,
the situation is reversed when investigating dynamical laser properties such as the switch-on process.
The time evolution of the output power is shown in Fig.~\ref{switch}. For better visibility, only 
the first 300 ps are displayed, whereas the simulation is carried out until the steady state is reached.
The realization of a spatially more spread out carrier distribution for alignment B is associated with a small delay of about 10ps in the onset of lasing.
However, the optimized structure reaches the maximum intensity significantly faster, i.e., the switch-on process is improved. 
A comparison of the results to laser rate equation models \cite{Coldren:95} indicates that the optimized structure should be close to critical damping of the relaxation oscillations, \cite{Lorke:11} as only half an oscillation cycle is visible.
This finding suggests an improved modulation capability of the optimized structures compared to those currently realized experimentally.
\noteblue{We would like to point out the main difference stems not from the inclusion of z-dependent transport, 
but from a the modification of the energetic tuning between QDs and injector QW, that leads to more efficient carrier
capture into the QD ground state. This effect results in a more efficient provision of excited charge carriers for the laser transition and hence to more efficient modulation
properties.}

\noteblue{
In our calculations, we use a low inhomogeneous broadening of 15 meV, which corresponds to the currently best possible results. Higher inhomogeneous broadening reduces the number of QD that efficiently couple into the laser mode. For a non-optimal alignment, higher inhomogeneous broadening increases the fraction of QDs with better carrier capture. However, as these will be detuned from the laser mode positioned at the inhomogeneous distribution peak, the first effect mentioned still leads to a deterioration the laser emission properties.
}

In a recent paper, \cite{khanonkin2021principle} a TI-QD laser with alignment A was compared to a non-TI laser with nominally identical QDs. The experimental results demonstrate lower threshold currents for the TI-QD laser but also significantly lower bandwidth compared to the non-TI QD laser. Although this is a different comparison, it corroborates
that TI-QD lasers with alignment A have beneficial threshold but unfavorable modulation properties, as only parts of the QD ensemble experiences optimal coupling to the injector QW.

\noteblue{In conclusion, we have shown that the modulation performance of TI-based QD laser structures can be substantially improved by 
optimizing the energetic alignment between QD ensemble and injector QW. Faster carrier scattering rates result in
a more efficient carrier transport into the lasing ground-state and shorten the switch-on time. Simultaneously this leads to a concentration of carriers in the central QD layers of the device, which impacts the switch-on current.}

\section*{Supplementary Material}
The supplementary material contains a detailed description of the theoretical methods used to calculate the carrier scattering rates.

\section{Acknowledgment}
M.L and F.J acknowledge support from the German Science Foundation (DFG) under grant number JA619/16-2 and CPU time at HLRN (Berlin/G\"ottingen). This work was partially supported by The Israel Science Foundation under grant number 460/21. I.K. acknowledges the support of the Rothschild Fellowship (The Yad Hanadiv Foundation), and of the Helen Diller Quantum Center at Technion.

 \bibliography{mybib2}
 \bibliographystyle{unsrtnat}

\end{document}

\maketitle

\section*{Theory of carrier dynamics in tunnel injection structures\label{theocarrdyn}}

To determine the carrier scattering rates, we utilize a treatment of the carrier-phonon interaction beyond perturbation theory~\cite{Inoshita:97,Kral:98,Seebeck:05}, which has been introduced using the nonequilibrium Green's functions (GF) technique. From the Dyson equation a quantum kinetic equation can be derived \cite{Haug_Koch:04},  
\begin{equation}\label{eq:qk-scat}
\begin{split}
\frac{\partial f_{\alpha}(t)}{\partial t} = 2\text{ Re}\sum_{\beta}\int\limits^t_{-\infty}dt'~|M_{\alpha\beta}|^2
 &G^{ret}_{\beta}(t-t')\left[G_{\alpha}^{ret}(t-t')\right]^* \\
 \ast\, \Big\{\left[f_{\beta}(t')(1-f_{\alpha}(t'))\right] i\hbar \Big[&(1+n_\text{LO})e^{- i\omega_\text{LO}(t-t')}
 +n_\text{LO}e^{+ i\omega_\text{LO}(t-t')}\Big] \\
 -\left[f_{\alpha}(t')(1-f_{\beta}(t'))\right]  i\hbar \Big[&(1+n_\text{LO})e^{+ i\omega_\text{LO}(t-t')}
 +n_\text{LO}e^{- i\omega_\text{LO}(t-t')}\Big]\Big\}~.
\end{split}
\end{equation}
Here, $f_{\alpha}$ is the carrier occupation function of state $\alpha$, $n_\text{LO}$ is the phonon occupation, $\epsilon_{\alpha}$ is the energy of the state $\alpha$, and $\omega_{\text{LO}}$ is the phonon frequency. The matrix elements $M_{\beta\alpha}$ of the carrier-phonon interaction are calculated from $k\cdot p$ wavefunctions as obtained from the nextnano$^3$ software package \cite{nextnano3} \note{via
\begin{equation}\label{eq:mat}
\begin{split}
M_{\beta\alpha}&=\sum\limits_{\vec{q}}\frac{M_{LO}}{|\vec{q}|}|\braket{\phi_\alpha|e^{i\vec{q}\vec{r}}|\phi_\beta}|^2~.
\end{split}
\end{equation}
}
\note{The indices $\alpha$ and $\beta$ label all states of the QD, injector QW, and bulk reservoir.} 
For the retarded polaron Greens function $G_{\alpha}^{ret}$, the Kadanoff-Baym equations \cite{Seebeck:05} can be used to obtain the following equation of motion:  
\begin{equation}
   \Big[ i\hbar\frac{\partial}{\partial t} - \epsilon_{\alpha} \Big]~G^{\text{ret}}_{\alpha}(t) =  \delta(t) + \int\!dt' ~~ \Sigma^{\text{ret}}_{\alpha}(t-t') ~ G^{\text{ret}}_{\alpha}(t')~.
\label{eq:Dyson_G_ret}
\end{equation}
In random-phase-approximation (RPA) \cite{Mahan:90}, the retarded selfenergy is given by
\begin{equation}
      \Sigma^{\text{ret}}_{\alpha}(t) ~ = ~ i \hbar \sum_{\beta} ~|M_{\alpha\beta}|^2
      G^{\text{ret}}_{\beta}(t) ~ d^<(-t),
\label{eq:Sigma_ret}
\end{equation}
where the phonon propagators $d^\gtrless$ contain the phonon frequency $\omega_{\text{LO}}$ and the phonon population function $n_{\text{LO}}$ according to
\begin{equation}\label{eq:phonon_propagator}
   i\hbar ~d^\gtrless(t) = n_{\text{LO}} ~e^{\pm i\omega_{\text{LO}} (t)} + (n_{\text{LO}}+1) ~e^{\mp i\omega_{\text{LO}} (t)}
\end{equation}
\note{assuming dispersionless LO phonons. It should be noted that Eqs~\eqref{eq:Dyson_G_ret}-\eqref{eq:phonon_propagator} describe renormalization effects of the electronic states due to carrier-phonon interaction. Via the retarded polaron Greens function this renormalization is included in the quantum kinetic equation \eqref{eq:qk-scat}. The latter generalizes the calculation of carrier scattering via the Boltzmann-equation. Instead of strict energy conservation in terms of free-carrier energies, Eq.~\eqref{eq:qk-scat} contains a time integral over retarded Greens functions with interacting electronic states and phonon propagators.}

Quasi-particles as eigenstates of the interacting system are characterized by new energies in comparison to the free-particle energies. In the non-perturbative regime, the new energies represent a polaron shift as well as the dressing of the electronic states with a series of phonon replica due to emission and absorption processes. The broadening reflects the finite lifetime of the interacting states, caused by emission and reabsorption of phonons. It is the overlap of the energy spectra of these dressed states and their quasi-particle broadening, which is lifting the phonon resonance condition.

\note{In the framework of the Markov approximation, Eq.~\eqref{eq:qk-scat} can be rewritten as 
\begin{equation}\label{eq:scattering1}
  \frac{d}{dt}f_\alpha(t)=\sum\limits_\beta \left[1-f_\alpha(t)\right] \; \Gamma^{in}_{\alpha\beta}(t) -f_\alpha(t) \; \Gamma^{out}_{\alpha\beta}(t)~,
\end{equation}
using the in- and out-scattering rates,
\begin{equation}
\begin{split}
    \Gamma^{in}_{\alpha\beta}(t)=f_\beta(t) \; |M_{\alpha\beta}|^2 \int\limits^t_{-\infty}dt'&i\hbar~G^{ret}_{\beta}(t-t')\left[G_{\alpha}^{ret}(t-t')\right]^*\\&\Big[ (1+n_\text{LO})e^{- i\omega_\text{LO}(t-t')}
 +n_\text{LO}e^{+ i\omega_\text{LO}(t-t')}\Big]~,
 \end{split}
\end{equation}
\begin{equation}
\begin{split}
\Gamma^{out}_{\alpha\beta}(t)= \left[1-f_{\beta}(t)\right] \; |M_{\alpha\beta}|^2 \int\limits^t_{-\infty}dt' &i\hbar~G^{ret}_{\beta}(t-t')\left[G_{\alpha}^{ret}(t-t')\right]^* \\ &\Big[(1+n_\text{LO})e^{+ i\omega_\text{LO}(t-t')}
 +n_\text{LO}e^{- i\omega_\text{LO}(t-t')}\Big]~.
 \end{split}
\end{equation}
Assuming that the phonons are in thermodynamic equilibrium, $n_{\text{LO}}$ is given by a Bose-Einstein function with the LO-phonon frequency and the lattice temperature. Then the time integrals can be determined independent of the dynamical calculation and serve as generalizations of the $\delta$-function in the Boltzmann equation describing quasi-particle effects. }

\begin{table}[!h]
\begin{center}
\begin{tabular}{|c|c|c|}
  \hline
  scattering time        & initial state  & final state  \\
  \hline
  $\tau^i_\text{tun}$    & injector QW    & i-th QD      \\
  $\tau^i_\text{tun,b}$  & i-th QD        & injector QW  \\
  $\tau_\text{cap,QW}$   & 3d bulk        & injector QW  \\
  $\tau_\text{esc,QW}$   & injector QW    & 3d bulk      \\
  $\tau^i_\text{cap}$    & 3d bulk        & i-th QD      \\
  $\tau^i_\text{esc}$    & i-th QD        & 3d bulk      \\
  \hline
\end{tabular}
\end{center}
\caption{Scattering times for different capture and escape processes.}
\label{table_1}
\end{table}

\note{To determine the scattering times of the various processes listed in Table~1, the kinetic equation \eqref{eq:scattering1} is solved numerically in time. An initial state occupied according to a specified excitation density and an empty final state are selected as the initial condition. The time evolution of the occupation in the final state is fitted by a function 
\begin{equation}\label{eq:fit}
    n(t)=n_0\left( 1-\exp(-t/\tau) \right)~,
\end{equation}
where $\tau$ refers to the corresponding scattering time in the left column of Table~1. For the capture of carriers from the injector QW into a QD state, described by $\tau^i_\text{tun}$, and the capture of carriers from the 3d bulk states into a QD state, described by $\tau^i_\text{cap}$, we follow the QD population according to Eq.~\eqref{eq:fit}. For the scattering into a continuum of states, like the back scattering from a QD state into injector QW states with $\tau^i_\text{tun,b}$ or the back scattering from the injector QW into the 3d bulk states with $\tau_\text{esc,QW}$, we proceed accordingly with a time evolution starting with empty final states and intial states populated according to a given excitation density. In this case, the time evolution of the particle density in the final states is used to calculate the scattering time. On this basis, a set of excitation-dependent scattering times is created.}

In conventional QD laser structures, the Coulomb interaction provides another mechanism for carrier scattering that is equally important to the interaction of carriers with LO-phonons \cite{chow2011will}.
It is a peculiarity of the TI-QD design that here the scattering due to LO-phonons clearly prevails. This is due to the fact that, using the nomenclature of Ref.~\cite{Nielsen:04}, the Coulomb scattering processes are of ''capture'' type, which are known to have a lower efficiency. Accordingly, we have not included them in this paper.

\section*{Material parameters}
\begin{table}[!h]
\begin{center}
\begin{tabular}{|c|c|}
\hline
Parameter & Value\\
  \hline
  $\tau_\text{sp}$    & 1 ns  \\
 $\tau_\text{res}$    & 1 ns  \\
   $\tau_\text{QW}$    & 1 ns  \\
 $\gamma_h$  & 15meV     \\
 $\varepsilon_b$  & 12.5     \\
 $\hbar\omega_\text{LO}$  & 36meV     \\
  $D_N$   &   117 $\frac{\text{cm}^2}{s}$      \\
  $\mu_N$   &   $\frac{e}{kT}D_N$    \\
  $D_\text{QW}$    &   2$\times$10$^{13}$ /cm$^2$       \\
  $D_\text{res}$    &  3$\times$10$^{20}$ /cm$^3$      \\
  \hline
\end{tabular}
\end{center}
\caption{Time constants and other material properties used in the calculations. }
\label{table_1}
\end{table}

 \bibliography{mybib2}
 \bibliographystyle{unsrtnat}